\documentclass[a4paper,11pt]{article} \usepackage{jcappub}
\pdfoutput=1 

\usepackage{graphicx}
\usepackage{natbib}

\title{Efficient propagation of systematic uncertainties from calibration to analysis with the SnowStorm method in IceCube}

\author[15]{M. G. Aartsen,}
\author[53]{M. Ackermann,}
\author[15]{J. Adams,}
\author[11]{J. A. Aguilar,}
\author[19]{M. Ahlers,}
\author[25]{C. Alispach,}
\author[3]{B. Al Atoum,}
\author[36]{K. Andeen,}
\author[50]{T. Anderson,}
\author[11]{I. Ansseau,}
% ...
\affiliation[3]{Dept. of Physics, University of Texas at Arlington, 502 Yates St., Science Hall Rm 108, Box 19059, Arlington, TX 76019, USA}
% ...
\affiliation[11]{Universit{\'e} Libre de Bruxelles, Science Faculty CP230, B-1050 Brussels, Belgium}
% ...
\affiliation[15]{Dept. of Physics and Astronomy, University of Canterbury, Private Bag 4800, Christchurch, New Zealand}
% ...
\affiliation[19]{Niels Bohr Institute, University of Copenhagen, DK-2100 Copenhagen, Denmark}
% ...
\affiliation[25]{D{\'e}partement de physique nucl{\'e}aire et corpusculaire, Universit{\'e} de Gen{\`e}ve, CH-1211 Gen{\`e}ve, Switzerland}
% ...
\affiliation[36]{Department of Physics, Marquette University, Milwaukee, WI, 53201, USA}
% ...
\affiliation[50]{Dept. of Physics, Pennsylvania State University, University Park, PA 16802, USA}
% ...
\affiliation[53]{DESY, D-15738 Zeuthen, Germany}
% ...

\emailAdd{analysis@icecube.wisc.edu}

\abstract{Efficient treatment of systematic uncertainties that depend on a large number of nuisance parameters is a persistent difficulty in particle physics experiments. Where low-level effects are not amenable to simple parameterization or re-weighting, analyses often rely on discrete simulation sets to quantify the effects of nuisance parameters on key analysis observables. Such methods may become computationally untenable for analyses requiring high statistics Monte Carlo with a large number of nuisance degrees of freedom, especially in cases where these degrees of freedom parameterize the shape of a continuous distribution. In this paper we present a method for treating systematic uncertainties in a computationally efficient and comprehensive manner using a single simulation set with multiple and continuously varied nuisance parameters. This method is demonstrated for the case of the depth-dependent effective dust distribution within the IceCube Neutrino Telescope.}


\usepackage{lineno}
\begin{document}
\maketitle

\section{Introduction}

A pervasive challenge in experimental particle physics is the development
of robust treatments for systematic uncertainties. There is no
single correct approach to treating systematic uncertainties in any analysis, but
it is nevertheless recognized that a credible parameterization of ignorance about the experiment being undertaken
is a critical ingredient in any measurement. Furthermore, for analyses yielding unified confidence intervals or Bayesian intervals
in multi-dimensional spaces, neither an overly tight nor an overly
loose parameterization of ignorance can generally be considered to
be conservative - either may enhance the likelihood of a spurious
signal or overly strong exclusion under certain circumstances, and so it is vital to parameterize ignorance as accurately
as possible.   Given that no
parameterization of uncertainty is exact, and that un-modelled effects and simplifying assumptions will always be present, treatment of
systematic uncertainties is be a frequent source of difficulty in particle physics and beyond.

The goal of this work is to present a method used at the IceCube South Pole Neutrino Observatory~\cite{Halzen:2010yj,Aartsen:2016nxy} to propagate a particularly complex form of systematic
uncertainty from calibration constraints to analyses of atmospheric neutrinos, including searches for sterile neutrinos~\cite{TheIceCube:2016oqi}. The uncertainty
in question is that of the effective dust concentration as a function of depth within the ice of the IceCube neutrino telescope. For the purpose of this work, the term ``dust'' will be used to refer to all forms of optical impurity within the ice, although mineral dust itself is only one major component.

The dust distribution within IceCube and the methods used to characterize
it have been described at length elsewhere~\cite{Refworks:1} and ongoing campaigns
continue to provide deeper understanding of its optical properties~\cite{Askebjer:1997ep,Askebjer:1994yn,Price:97,GRL:GRL20535,AnisotropyPaper,JGRD:JGRD12654}. In past generations of IceCube analyses, the detailed properties
of the bulk ice in the array could be shown to be a sub-dominant concern (for example, in Ref~\cite{Jones:2015bya}),
with their effect on analysis distributions being comfortably less
than statistical uncertainty in most samples. The latest generation
of atmospheric neutrino analyses at IceCube, on the other hand, consider
samples containing more than 250,000 muon neutrino candidates~\cite{Jones:2019nix}. For
these analyses, percent-scale precision
on event distributions is required. This mandates a detailed understanding
of the effect of uncertainty on ice properties such as dust distribution
within analyses. 

The reasons why the dust distribution in IceCube is a particularly
challenging uncertainty to parameterize include the following:
\begin{itemize}
\item The effect of the dust concentration vs depth on variables such as
reconstructed muon energy and zenith angle is highly non-trivial to
estimate. The only known way to establish this effect is through Monte
Carlo simulation of an event within a given ice model, and it is computationally
unfeasible to generate Monte Carlo samples given every reasonable
ice model configuration.
\item The dust concentration varies continuously as a function of depth.
The uncertainty on this continuous function must be constrained and
propagated without introducing an overwhelming number of nuisance parameters.
\end{itemize}
These two challenges are not unique to this source of uncertainty
or to IceCube, and the method presented here may be used to treat 
uncertainties on continuous functions in highly multidimensional spaces
in other contexts. The organization of this paper is such that we
treat these two challenges in order. The first challenge is addressed
in Sec.~\ref{sec:SnowStorm}, via construction of the SnowStorm method, a Monte Carlo
technique whereby constraints on many nuisance variables can be mapped
from calibration data to analysis space in a computationally efficient
manner. The second challenge is then addressed in Sec.~\ref{sec:Ice}, focusing on
the specific example of the effective dust distribution in IceCube. A Fourier
parameterization is developed that allows for propagation of the most
important uncertainty contributions from calibration to analysis space,
and the SnowStorm approach is applied to generate a covariance
matrix for use in physics analysis.

\section{The SnowStorm MC Method\label{sec:SnowStorm}}

\subsection{Measurements and systematic uncertainties\label{sec:MeasUncert}}

A wide class of measurements in particle physics are made by comparing
some observed distribution of events $\phi(E_{\alpha})$ - for example,
a measured neutrino spectrum as a function of energy - against a predicted
distribution $\Psi_{\vec{\rho}}(E_{\alpha})$, generally derived from calculations
and Monte Carlo simulations, informed by calibration data. The distribution to be measured is a function of some variable $E_\alpha$, where $\alpha$ may be either a simple index (for example, labelling energy bin number), a compound index (for example, energy and zenith bin numbers), or a continuous variable (in a non-binned analysis).  The prediction
depends on some ``physics parameters'' to be measured $\vec{\rho}$,
and carries some uncertainty on associated with systematic and statistical
effects. In the regime where Gaussian propagation of errors is valid,
the uncertainty on $\Psi_{\vec{\rho}}(E_{\alpha})$ can be completely described as a covariance matrix
in the analysis space, $\Sigma_{\vec{\rho}}(E_{\alpha},E_{\beta})$. A measurement
of $\vec{\rho}$ is made by comparing $\phi(E_{\alpha})$ with $\Psi_{\vec{\rho}}(E_{\alpha})$
given $\Sigma_{\vec{\rho}}(E_{\alpha},E_{\beta})$ at each $\vec{\rho}$. Exactly
what the outcome of this measurement represents and how the final
result is constructed depends on whether a frequentist or Bayesian
probability construction is preferred, and we will not review this
topic here. What is important is that in either case, the proper estimation
of $\Sigma_{\vec{\rho}}(E_{\alpha},E_{\beta})$ is central to calculation of the final result.

Any effect that influences the analysis distribution and that depends
on some parameters that are imperfectly known will contribute to $\Sigma_{\vec{\rho}}(E_{\alpha},E_{\beta})$.
The unknown parameters on which this effect depends are typically called ``nuisance parameters''
and we represent them as a vector $\vec{\eta}$. Given some choice of $\vec{\eta}$,
the predicted event distribution can be calculated, and we denote
this as $\psi_{\vec{\rho},\vec{\eta}}(E_{\alpha})$. Choosing coordinates
such that the ``central'' prediction is one corresponding to $\vec{\eta}=\vec{0}$,
we impose the definition
$\Psi_{\vec{\rho}}(E_{\alpha})\equiv\psi_{\vec{\rho},\vec{0}}(E_{\alpha})$. 

Broadly speaking, there are two typical things an analyzer may want
to know about the effects of the nuisance parameters on the prediction:
\begin{enumerate}
\item What is the effect of varying $\vec{\eta}$ on the predicted distribution
of events? In other words: how does $\psi_{\vec{\rho},\vec{\eta}}(E_{\alpha})$
depend on $\vec{\eta}$?
\item Given some set of constraints on $\vec{\eta}$ from calibration data
or other experiments, what is the uncertainty on $\Psi_{\vec{\rho}}(E_{\alpha})$?
In other words, given available knowledge of $\vec{\eta}$, what is
$\Sigma_{\vec{\rho}}(E_{\alpha},E_{\beta})$?
\end{enumerate}
For some nuisance parameters the dependence of $\psi_{\vec{\rho},\vec{\eta}}(E_{\alpha})$ upon $\eta_i$ is straightforward to model, and techniques such as re-weighting can be used to answer the first question. However, for other classes of systematic uncertainty, for example, effects that change detector response in non-trivial ways, the effect on final analysis-level distributions is highly non-trivial to estimate.  A common approach for these classes of uncertainty involves production of many distinct Monte Carlo sets with different choices of nuisance parameters.  This approach, sometimes described as a ``multiple Universes'' scheme, becomes quickly
computationally untenable as the dimensionality of $\vec{\eta}$ becomes
large, especially for high-statistics samples where the corresponding
Monte Carlo samples must have high event counts.  It can also lead to neglecting of important correlations between different $\eta_i$ variables. 

For this reason and others, enumeration of a suitable suite of $\vec{\eta}$ parameters and answering of
the two questions posed above is often the most time-consuming aspect of a carefully constructed particle physics analysis.  The SnowStorm Monte Carlo method described in this paper is a recipe for answering
both questions (1) and (2), given an arbitrarily large number of $\vec{\eta}$
parameters using a single Monte Carlo ensemble.

\subsection{The SnowStorm Monte Carlo ensemble}

\begin{figure}[t]
\centering
\includegraphics[width=0.49\textwidth]{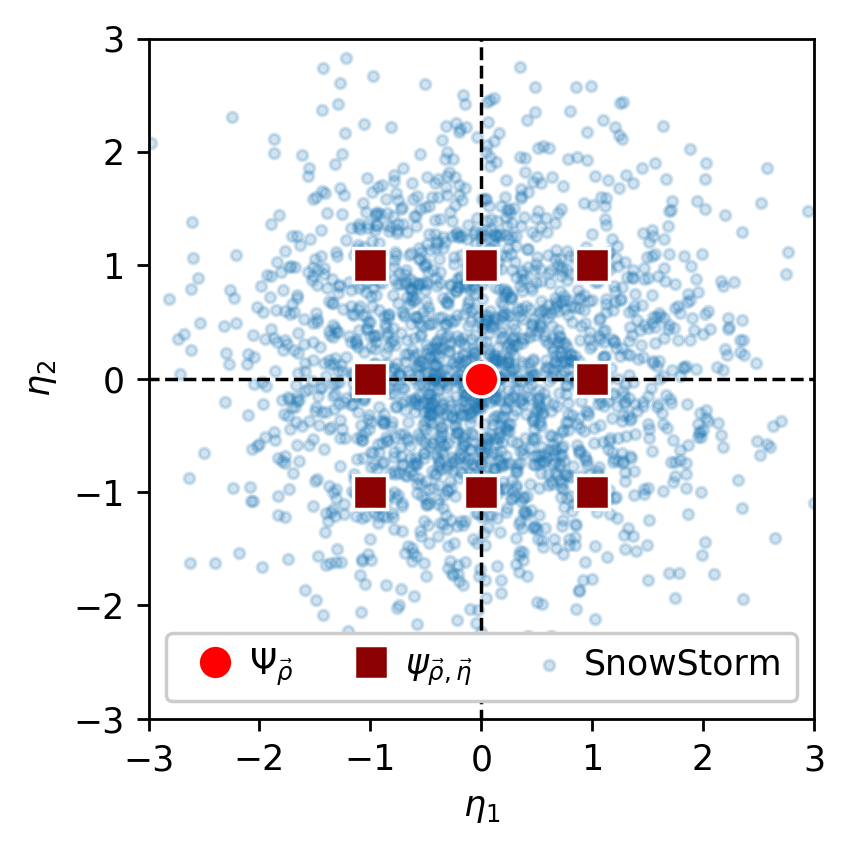}
\includegraphics[width=0.49\textwidth]{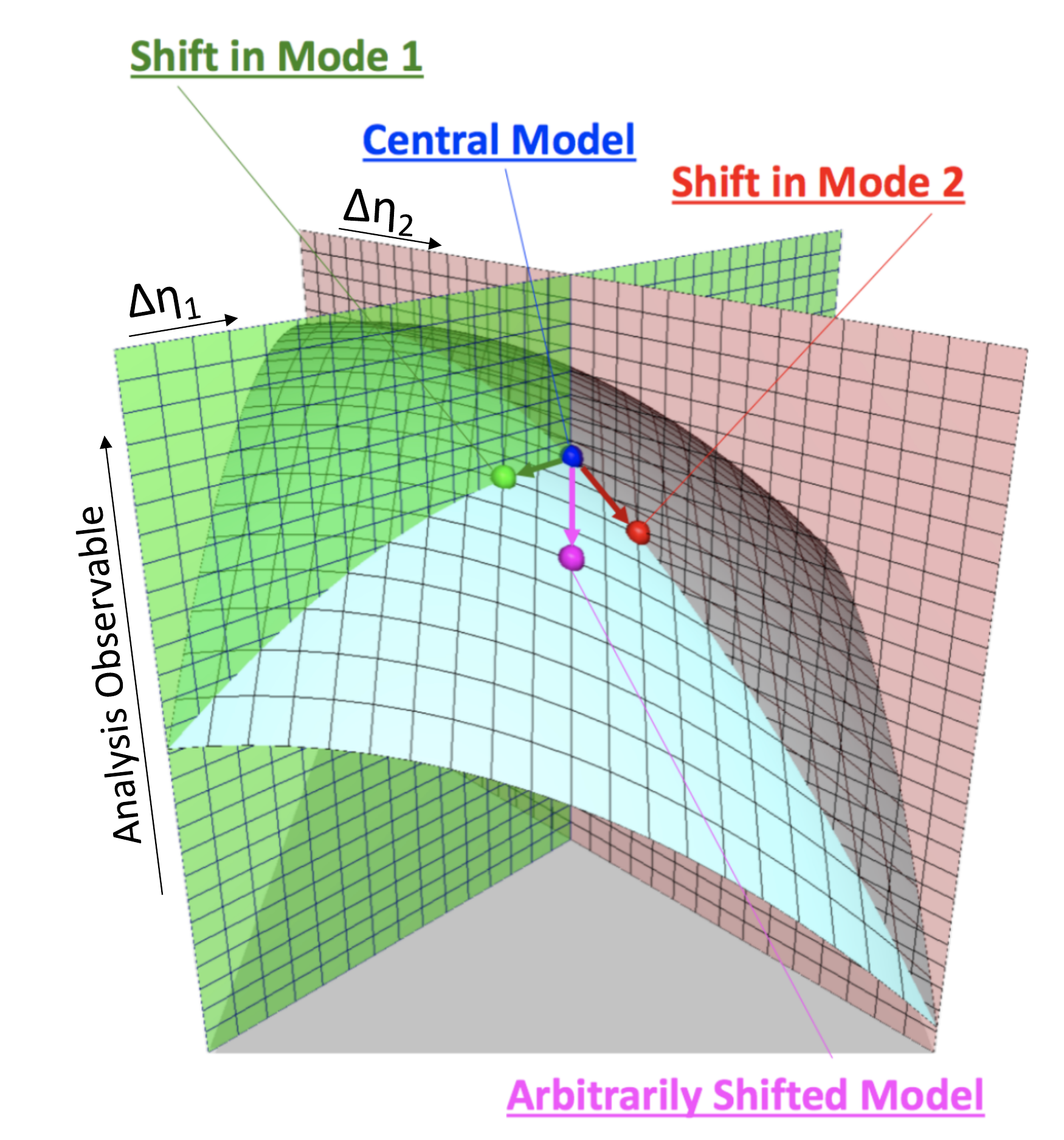}
\caption{Left: Cartoon illustrating the difference between many discrete shifts in nuisance parameters (red squares) -- each requiring an entire Monte Carlo set -- compared with a single SnowStorm Monte Carlo (small blue dots). Right: the method of extracting independent gradients and building up a covariance matrix is accurate in the limit that the analysis space is locally flat as the nuisance parameters are varied. This corresponds to linear perturbativity.\label{fig:MultiSimCartoon}}
\end{figure}

A SnowStorm Monte Carlo ensemble\footnote{The name SnowStorm reflects the idea that each event in the sample is distinct, like each snowflake in a snowstorm} consists of a single simulation set where every event (or every group of a few events) is generated with a unique set of nuisance parameters $\vec{\eta}$.  Under some loose assumptions, given a suitably prepared SnowStorm Monte Carlo sample, it is possible to assess the effects of these variations on an analysis by appropriately manipulating this ensemble. The structure of the SnowStorm ensemble in $\vec{\eta}$ space is compared schematically to a ``multiple Universes'' based approach in Fig.~\ref{fig:MultiSimCartoon}, left.

An assumption underlying the SnowStorm method is that the effects of systematic uncertainties on analysis variables are sufficiently small that they can be treated perturbatively.  Such an assumption also underlies, explicitly or implicitly, the majority of methods for systematic uncertainty estimation and error analysis in common use.  We also assume that the statistical uncertainty on Monte Carlo event counts is very small compared to that on the data.  Under such conditions, summation over a Monte Carlo ensemble with different values of $\vec{\eta}$ distributed according to a distribution function $P(\vec{\eta})$  can be replaced by an integration:
\begin{equation}
\sum_{all \, \vec{\eta}}\rightarrow \int d^N\eta P(\vec{\eta}).
\end{equation}
The distribution function $P(\vec{\eta})$ is specified when generating the Monte Carlo ensemble.  We enforce on $P(\vec{\eta})$ two properties: it is normalized and
it is symmetric in every variable $\eta_{i}$ such that:
\begin{equation}
P(\eta_{i},\vec{\eta}_{j\neq i})=P(-\eta_{i},\vec{\eta}_{j\neq i})\quad\forall i,
\end{equation}
\begin{equation}
\int d^{N}\eta\:P(\vec{\eta})=1.
\end{equation}
Much of what follows conceptually relies only on these assumptions. For simplicity, however, we make an explicit choice of $P(\vec{\eta})$ choosing a product of Gaussian functions of equal width $\sigma$:
\begin{equation}
P(\vec{\eta})=\prod_{i}\frac{1}{\sigma\sqrt{2\pi}}e^{-\eta_{i}^{2}/2\sigma^{2}}.
\end{equation}
The equal-width requirement corresponds to a choice of coordinates for $\vec{\eta}$, rather than a constraint on the size of any physical effects to be modelled.

If the effects of the nuisance parameters $\vec{\eta}$ are perturbative,
this implies that the distribution function at any $\vec{\eta}$ can be written as a Taylor expansion around the central distribution:
\begin{equation}
\psi_{\vec{\rho},\vec{\eta}}=\Psi_{\vec{\rho}}+\vec{\eta}.\vec{\nabla}_{\eta}\left[\psi_{\vec{\rho},\vec{\eta}}\right]_{\vec{\eta}=\vec{0}}+{\cal O}(\eta^{2})\label{eq:expansion},
\end{equation}
with the largest effects present in the first term. Neglecting the
${\cal O}(\eta^{2})$ terms implies making a locally Euclidian approximation
to the nuisance space as illustrated in Fig~\ref{fig:MultiSimCartoon}, right. This is
appropriate for a wide class of effects, and always true for sufficiently
small perturbations in $\vec{\eta}$. This assumption should
not be taken for granted, however.  Its validity can be tested given
availability of both a central Monte Carlo set $\Psi_{\vec{\rho}}$
and a SnowStorm ensemble, by considering the integrated prediction of the SnowStorm sample:
\begin{equation}
\psi_{\vec{\rho}}^{SnowStorm}=\int d\vec{\eta}P(\vec{\eta})\psi_{\vec{\rho},\vec{\eta}}=\int d\vec{\eta}P(\vec{\eta})\left[\Psi_{\vec{\rho}}+\vec{\eta}.\vec{\nabla}_{\eta}\left[\psi_{\vec{\rho},\vec{\eta}}\right]_{\vec{\eta}=\vec{0}}+{\cal O}(\eta^{2})\right].
\end{equation}
The integral over the distribution function in the first term normalizes
to 1 since $\int d\vec{\eta}P(\vec{\eta})=1$. The second term is
an odd function of $\vec{\eta}$ integrated between even limits and as such
gives no contribution; The final term encodes all the residual higher
order effects. Thus:
\begin{equation}
\psi_{\vec{\rho}}^{SnowStorm}=\Psi_{\vec{\rho}}+\int d\vec{\eta}P(\eta){\cal O}(\eta^{2}).\label{eq:TaylorSnow}
\end{equation}
The prediction of the SnowStorm ensemble and the central model are
identical up to ${\cal O}(\eta^{2})$ effects. If comparison shows
they are equivalent within the available statistical uncertainty,
it is appropriate to neglect the non-linear terms in $\vec{\eta}$ to this same precision.
Assuming this test passes and the
linear and perturbative approximation is a good one, Eq.~\ref{eq:expansion} reduces to:
\begin{equation}
\psi_{\vec{\rho},\vec{\eta}}=\Psi_{\vec{\rho}}+\vec{\eta}.\vec{G}_{\vec{\rho}},\quad\quad \vec{G}_{\vec{\rho}} \equiv \vec{\nabla}_{\eta}\left[\psi_{\vec{\rho},\vec{\eta}}\right]_{\vec{\eta}=\vec{0}}. \label{eq:expansion2}
\end{equation}
The prediction of any model in $\eta$ space can be obtained given $\Psi_{\vec{\rho}}$ and a list of $N$ ``nuissance gradient'' coefficients $\vec{G}_{\vec{\rho}}$. 

For analyses where the effects of the physics parameters are relatively small in the region of interest, it is possible to make a further approximation that $\vec{G}_{\vec{\rho}} =\vec{G}_{\vec{0}}$, since the relevant dependencies on $\vec{\rho}$ would be second order perturbative effects in Eq.~\ref{eq:expansion2}.  This reduces computational load to calculating only one universal set of $N$ gradients, rather than one set for each value of the physics parameters $\vec{\rho}$. This assumption is not a necessary condition for use of the SnowStorm method, however, and its suitability should be assessed on a per-case basis. 

\subsection{Nuisance gradient extraction from the SnowStorm ensemble}
Under the assumption of linearity and perturbativity (testable by comparing the SnowStorm ensemble to a central model), a method of extracting the $N$ nuisance gradients $G_i$ would provide an answer to Question (1) of Sec~\ref{sec:MeasUncert}: that is, the effect of making small changes in each of the nuisance parameters on the distribution of the analysis observable.  There are two approaches to extracting nuisance gradients, which will now be described.  In the limit of infinite statistical precision and perfect linearity, these two methods should provide identical results.  In subsequent analysis will use only the first approach, which offers better statistical precision on $\vec{G}$ for our case of interest.

First, consider making a cut of the SnowStorm ensemble along one direction
$\eta_{i}=0$. The sample is then divided into two sub-samples, $\psi_{\vec{\rho}}^{i+}$
and $\psi_{\vec{\rho}}^{i-}$, which each encode predictions:
\begin{eqnarray}
\psi_{\vec{\rho}}^{i+}=\int_{0}^{\infty}d\eta_{i}\int_{-\infty}^{\infty}d^{N-1}\eta P(\vec{\eta})\left[\Psi_{\vec{\rho}}+\vec{\eta}.\vec{\nabla}_{\eta}\left[\psi_{\vec{\rho},\vec{\eta}}\right]_{\vec{\eta}=\vec{0}}\right],
\\
\psi_{\vec{\rho}}^{i-}=\int_{-\infty}^{0}d\eta_{i}\int_{-\infty}^{\infty}d^{N-1}\eta P(\vec{\eta})\left[\Psi_{\vec{\rho}}+\vec{\eta}.\vec{\nabla}_{\eta}\left[\psi_{\vec{\rho},\vec{\eta}}\right]_{\vec{\eta}=\vec{0}}\right].
\end{eqnarray}
The $N-1$ integrals along directions other than $i$ give
no contribution from the linear term, by symmetry. However, symmetry of the integral along the $i$ direction is broken by the cut, and a finite contribution can be calculated by evaluating the integral
over $\eta_{i}P(\eta)$ in this half space, yielding:
\begin{equation}
\psi_{\vec{\rho}}^{i\pm}=\frac{1}{2}\Psi_{\vec{\rho}}\pm\frac{\sigma}{\sqrt{2\pi}}G_{\vec{\rho},i}.
\end{equation}
The gradient function $G_{\vec{\rho},i}$ is then straightforwardly
extracted by subtraction:
\begin{equation}
G_{\vec{\rho},i}=\frac{1}{\sigma}\sqrt{\frac{\pi}{2}}\left(\psi_{\vec{\rho}}^{i+}-\psi_{\vec{\rho}}^{i-}\right).
\end{equation}
Cutting the SnowStorm ensemble in each direction one by one, the $N$
dimensional vector $\vec{G}_{\vec{\rho}} \equiv \vec{\nabla}_{\eta}\left[\psi_{\vec{\rho},\vec{\eta}}\right]_{\vec{\eta}=\vec{0}}$ can be extracted completely from one Monte Carlo ensemble.

\begin{figure}[t]
\includegraphics[width=0.99\columnwidth ]{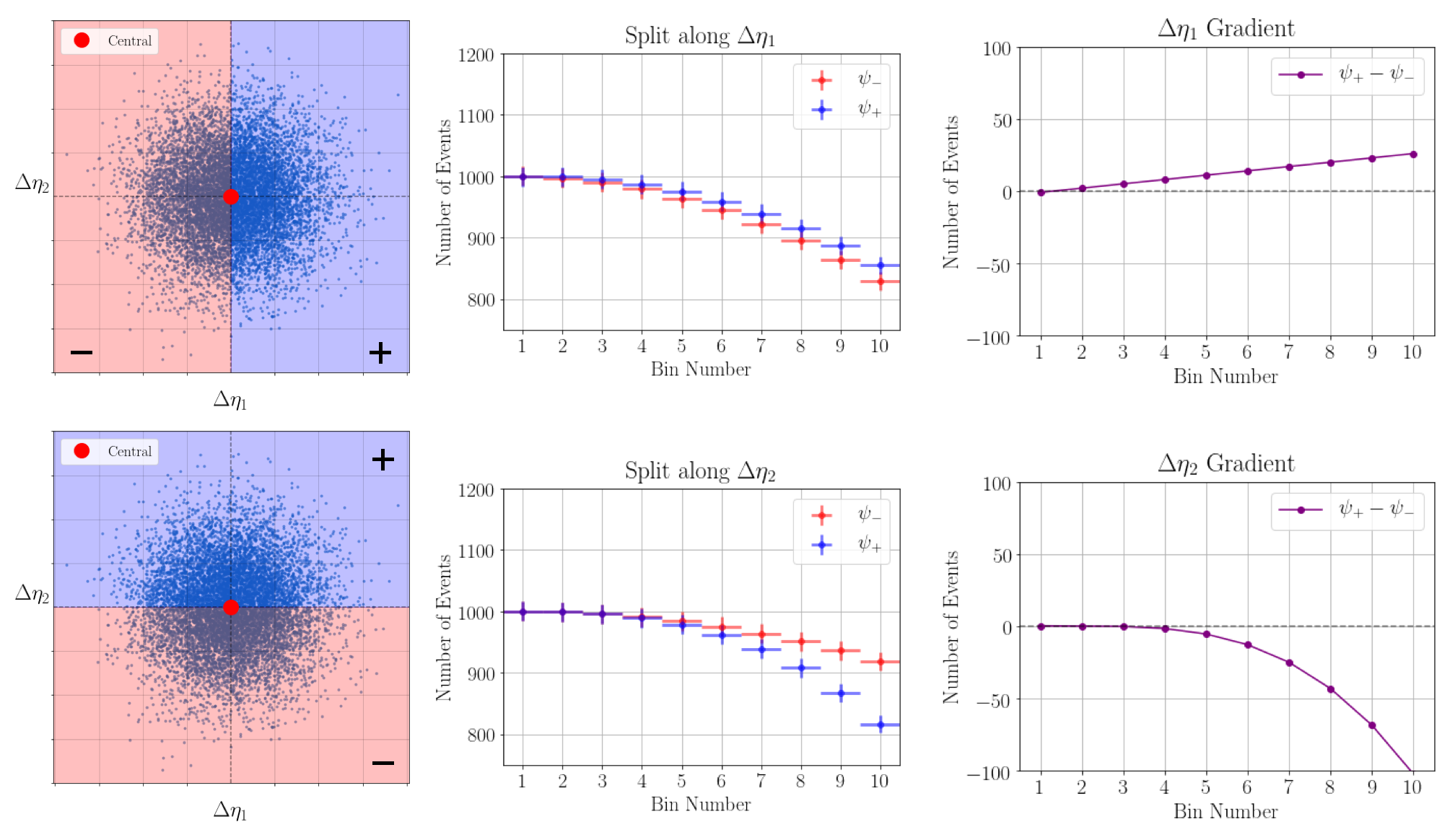}
\caption{Cartoon illustrating procedure for extracting gradients. The distribution is split along a given nuisance parameter direction into positively perturbed and negatively perturbed sets. The gradients are then proportional to the difference of the two sets.}
\label{fig:GradientCartoon}
\end{figure}

There is a second approach to extract $\vec{G}$, involving weighting
rather than cutting. In this case, consider constructing a prediction
where each event in the SnowStorm ensemble is weighted by a factor of $\eta_{i}$,
for one chosen direction in $\vec{\eta}$ space. This prediction takes the form:
\begin{equation}
\tilde{\psi}_{\vec{\rho}}^{i}=\int_{-\infty}^{\infty}d^{N}\eta\,\eta_{i}P(\vec{\eta})\left[\Psi_{\vec{\rho}}+\vec{\eta}.\vec{\nabla}_{\eta}\left[\psi_{\vec{\rho},\vec{\eta}}\right]_{\vec{\eta}=\vec{0}}\right].
\end{equation}
Now, the left term integrates to zero by symmetry, as do all the terms
in the dot product except the one for the $i$ component, which leaves
a residual contribution due to its even integrand:
\begin{equation}
\tilde{\psi}_{\vec{\rho}}^{i}=\frac{\sigma^{2}}{2}G_{\vec{\rho},i}.
\end{equation}
Which is trivially inverted to obtain a second, different expression
for the elements of $\vec{G}$. We thus see that either by weighting or cutting the sample, the full
nuisance gradient vector $\vec{G}$ can be extracted, and from this,
the prediction of any nearby model in $\vec{\eta}$ space can be obtained
via Eq.~\ref{eq:expansion2}.

Our experience has shown that somewhat better statistical
precision on $\vec{G}$ is obtained using the cutting method, since
the weighting method exaggerates the contribution of events in the
tails of the $P$ distribution, leading to larger fluctuations. A
schematic illustration of the cutting method applied on a two-dimensional
$\vec{\eta}$ space is shown in Fig.~\ref{fig:GradientCartoon}.
 
\subsection{Propagation of calibration constraints and nuisance priors to analysis space}

Implicitly, all distribution functions $\phi,\Psi,\psi$ and the gradients $\vec{G}$ are functions of the analysis variable $E_\alpha$, though we had suppressed this dependence for notational convenience. Here we restore it, as:
\begin{equation}
    G_i(E_\alpha)\equiv G_{i;\alpha}, \quad \psi_{\vec{\rho},\vec{\eta}}(E_\alpha)\equiv \psi_{\vec{\rho},\vec{\eta};\alpha},\quad \mathrm{etc...}
\end{equation}
In general, the nuisance parameters $\vec{\eta}$ will not be totally unknown, but subject to some constraints either from calibration data or from previous measurements.  These constrains can be represented in the form of a covariance matrix $\Xi$, which would be diagonal if all $\eta_i$ were independent, but generally will not be if the $\vec{\eta}$ values are constrained by common calibration data.  A basic theorem (proof reviewed in Appendix) can be invoked to map $\Xi$, the covariance matrix on the nuisance parameters $\vec{\eta}$, to $\Sigma$, the covariance matrix in the binned analysis space:
\begin{equation}
\Sigma_{\alpha,\beta}=G_{i;\alpha} \Xi_{ij} G_{j;\beta},\label{eq:TheTheorem}
\end{equation}
where $\vec{G}$ is the gradient vector (representing the effects of small shifts in the nuisance parameters on analysis-space distributions) derived in the previous section. Thus, given constraints on $\vec{\eta}$ which may or may not involve correlations, and the gradient vector $\vec{G}$, the analysis covariance $\Sigma_{\alpha,\beta}$ can be extracted, providing an answer to Question 2 of Sec~\ref{sec:MeasUncert}.  Exactly how the matrix $\Xi$ can be derived will vary by experiment and by effect. It is common to use analysis side-bands or calibration data for this purpose, to constrain the nuisance parameters in the sample.  An illustrative example where an independent calibration data set is used to construct $\Xi$ will be presented in the following section.
 
\section{The SnowStorm method applied to the IceCube dust distribution~\label{sec:Ice}}

\begin{figure}[t]
\centering
\includegraphics[width=\textwidth]{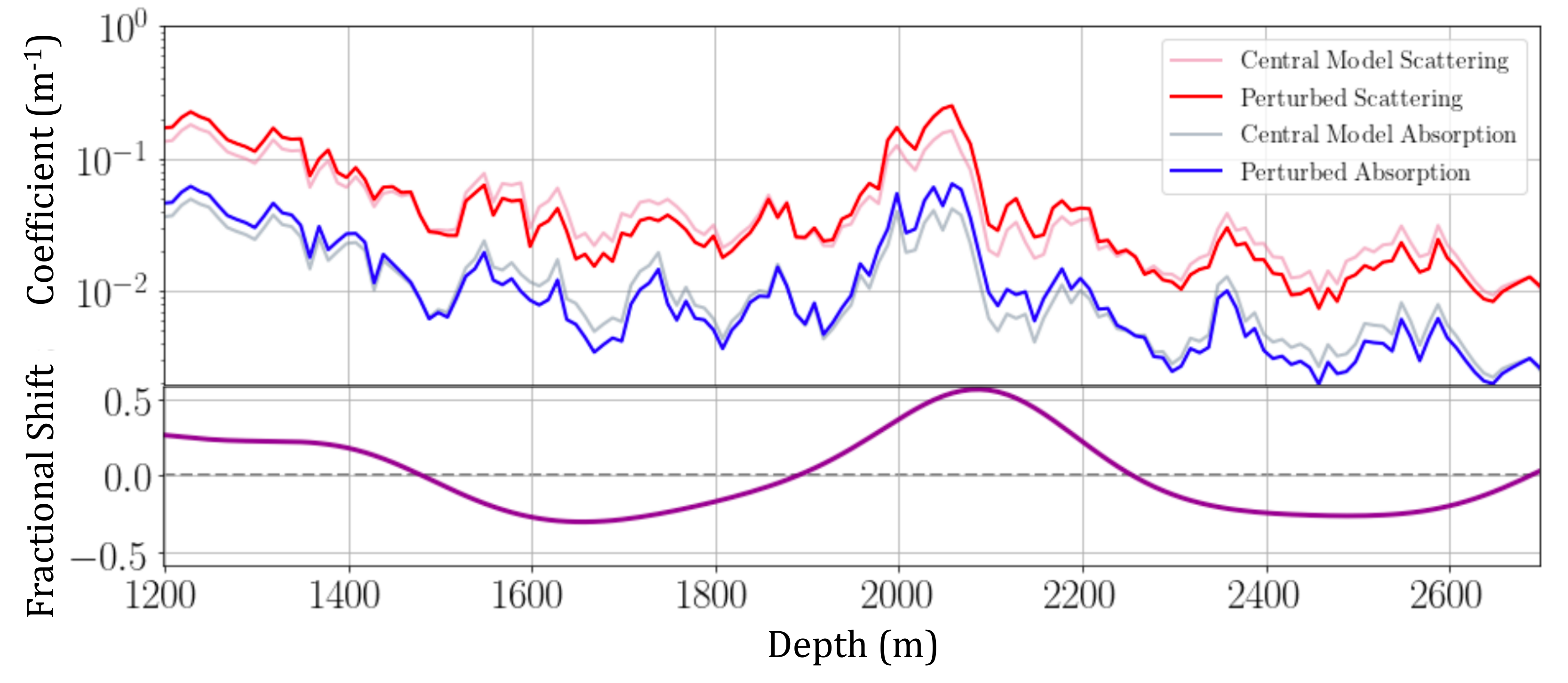}
\caption{An illustrative (but relatively large) perturbation of the ice parameters demonstrating depth dependant features. This example uses a relative shift of +50\% in amplitudes 2 and 5 and -50\% in amplitude~3.}
\label{fig:Illustration}
\end{figure}

The optical properties of the ice at the South Pole influence the propagation and detection of photons within the IceCube detector. It has long been established that the ice has properties that are depth dependent and correlate with the climate history on Earth over the period of its accumulation, of approximately 100,000 years~\cite{Icecube:2017vgw,GRL:GRL13620}.  The primary optical effects that influence photon propagation are absorption and scattering, and these are observed to exhibit a strong depth dependent correlation with one-another, suggesting a common origin.  This origin is generally understood to be small concentrations of dust throughout the array. Since the dust concentration is very low, the dust distribution is best probed via its effects on light propagation, using both dedicated dust-logger devices~\cite{GRL:GRL13620} and in-situ calibration using LED flashers within the IceCube array~\cite{Refworks:1}.  Based on this calibration data, a layered ice optical model has been constructed, with absorption and scattering coefficients assigned for each 10m layer, their values continuing to be refined as understanding of the ice properties improves through evolving calibration techniques and the incorporation of increasingly comprehensive catalogues of optical effects.  The structure and tuning of the ice model is described at length elsewhere, and a detailed exposition of its construction is outside the scope of this work.  The absorption and scattering coefficients from a recent generation of IceCube ice model are shown in Fig~\ref{fig:Illustration}, as well as an example of a perturbed model illustrating the continuous freedom in these parameters which must be explored to properly account for its uncertainty.

IceCube analyses typically rely on comparison of data to Monte Carlo, with this Monte Carlo ideally generated using the most up-to-date generation of ice model available.   As an illustration of the SnowStorm method we present its application to a specific form of ice uncertainty- that induced by imprecise knowledge depth-dependent distribution of dust inducing scattering and absorption within the IceCube array. We do not treat here the uncertainty on the form or composition of dust (i.e. the possibility that some areas of dust could create more absorption and less scattering, for example), uncertainty on the source and form of optical anisotropy~\cite{AnisotropyPaper}, uncertainty on the properties of refrozen ice in the vicinity of the Digital Optical Modules, or other possible sources of un-modeled uncertainty. The resulting uncertainty estimate should thus not be considered as a comprehensive uncertainty budget for all IceCube optical effects, or as a strong statement about glaciology.  Rather, we seek to construct a method for propagating a known, significant form of optical uncertainty for high-statistics atmospheric neutrino analyses in IceCube, that has been hitherto very difficult to assess.

\subsection{Model parameter reduction via Fourier analysis}

\begin{figure}[t]
\centering

\includegraphics[width=0.85\textwidth]{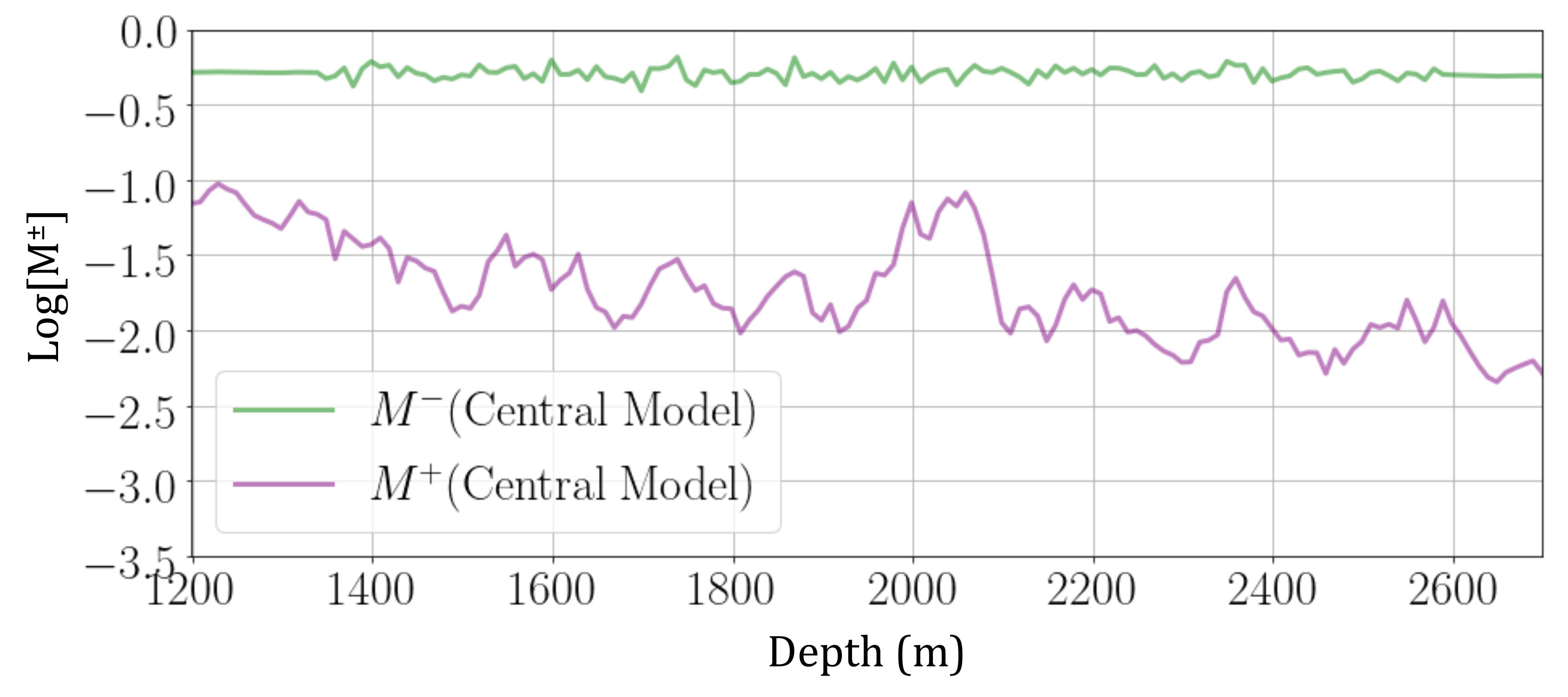}
\caption{The spectrum to be Fourier decomposed from central model for the re-parametrization. Note the flatness of $M^-$ indicating a strong correlation between scattering an absorption over all depths.}
\label{fig:M+M-}
\end{figure}

Due to the large number of parameters in the ice model combined with the substantial computational load required to accurately simulate light propagation within the ice, traditionally the most efficient means of assessing the effects of ice uncertainties on high-energy and high-statistics analyses has been to create discrete simulation sets where the absolute scales of bulk ice properties (e.g. scattering and absorption) are shifted by the same fraction across the entire detector. This method is fundamentally restricted in its ability to accurately assess depth-dependent contributions to systematic uncertainty.

One possibility would be to vary individual layers one by one, and consider the scale of these variations as entries in the $\vec{\eta}$ vector.  However, it is clear that varying only a single layer can have at most a very modest effect on the distribution of analysis variables such as reconstructed energy when integrated over the full sample of detected events. The ice model variants of most interest, which may strongly impact analysis outcomes, are those where correlated shifts in the ice model are made over certain larger regions of the detector.  These could be accounted for with a layer-by-layer variation protocol, given a suitably derived correlated covariance matrix for $\vec{\eta}$.  This would require many thousands of independent parameter constrains to be derived from flasher data, two for each layer and their covariances, and very strong local anti-correlations would be expected. 

\begin{figure}[t]
\centering
\includegraphics[width=\textwidth]{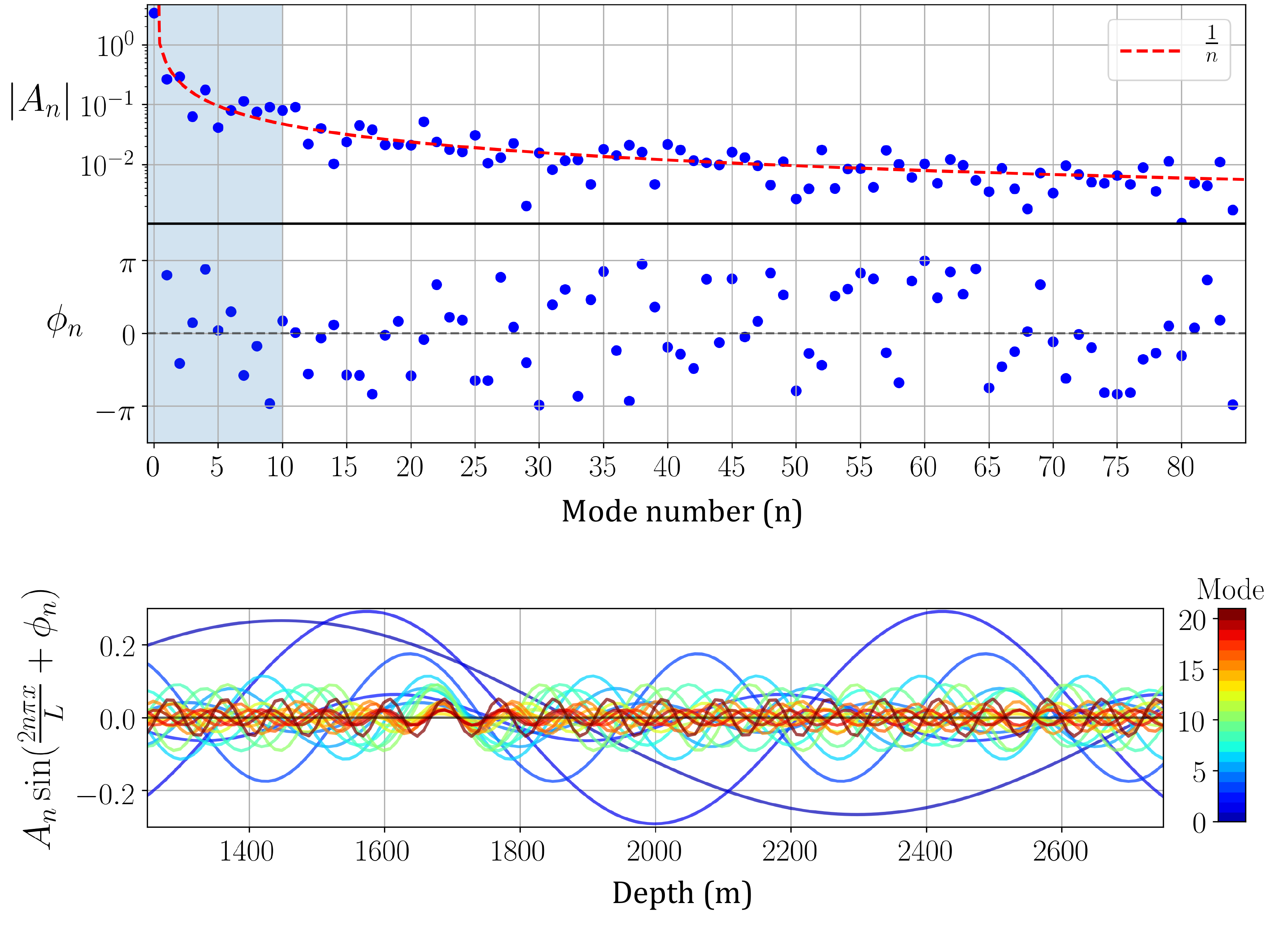}
\caption{Top: The parameters from the Fourier decomposition of the central model. The blue region highlights the leading modes considered in this work. Note the rate of convergence of the amplitudes goes as $~ 1/n$, consistent with a well-behaved series. Bottom: the depth dependant structure of each term in the Fourier expansion (with mode zero omitted).}
\label{fig:Params}
\end{figure}

To avoid generating constraints within the aforementioned large layer-by-layer nuisance covariance matrix, a significantly more efficient method has been developed.  Rather than describing the ice model layer-by-layer, it can equivalently be described in Fourier space, with entries in $\vec{\eta}$ being the amplitudes and phases of Fourier modes.  A fully covariant constraint on $\vec{\eta}$ derived in Fourier space would yield an equivalent uncertainty envelope to one derived in depth space. However, the benefit of Fourier space is that it offers a natural ordering principle to select the most important uncertainty contributions.  Low frequency Fourier modes have large amplitudes, and their variation reflects macroscopic changes over large regions of the IceCube detector.  These are expected to contribute the dominant parts of analysis-space uncertainties.  The high frequency modes, on the other hand, have smaller amplitudes and their variation is expected to have a minimal effect in analysis space, since the effect of the fast-varying ice properties effectively cancels when integrating over the detector and accounting for resolution effects. It will be demonstrated that this intuitive picture is valid in Sec.~\ref{sec:NuisCov}.

A notable feature of absorption and scattering coefficients shown in Fig.~\ref{fig:Illustration} is that they are highly correlated with each other as a function of depth.  These two distributions can thus be represented in terms of a completely correlated part, which contains most of the information about their shapes (physically, the effective dust concentration vs. depth), and a completely anti-correlated part, which encodes the far sub-leading effects that determine their differences (physically, changes in dust composition or shape throughout the array).  

To this end, we take the discrete Fourier decomposition of the sum ($M^+$) and difference ($M^-$) of the logarithms of the scattering and absorption coefficients at their central-model values.  These can be considered to represent the totally correlated and totally anti-correlated parts of the absorption and scattering profiles, respectively. The depth profiles of $M^+$ and $M^-$ are shown in Fig.~\ref{fig:M+M-}.

\begin{equation}
M^+(x) \equiv \frac{1}{2}\log_{10}\big({Abs \times Sca}\big) = \frac{A_0}{2} + \sum_n{A_n \sin\Big({\frac{2 n \pi x}{L}}+\phi_n\Big)},
\end{equation}
\begin{equation} 
M^-(x) \equiv \frac{1}{2}\log_{10}\big({Abs \div Sca}\big) = \frac{B_0}{2} + \sum_n{B_n \sin\Big({\frac{2 n \pi x}{L}} + \gamma_n\Big)}.
\end{equation}
From these variables, we may recover the physical parameters via the transformation:
\begin{equation}
Abs = 10^{M^+ + M^-}\quad\quad Sca = 10^{M^+ - M^-},
\end{equation}
with all of $Abs$, $Sca$, $M^+$ and $M^-$ being functions of depth, $x$. The zeroth Fourier mode of $M^-$ encodes the absolute relative scale of absorption vs. scattering, which is substantial, whereas the higher modes encode their depth-dependent shape differences, and are very small.  This can be seen in the flatness of $M^-$ in Fig.~\ref{fig:M+M-}, which reflects very little variation in the ratio of absorption to scattering across the detector within the best-fit ice model. The variations in $M+$ that increase or reduce absorption or scattering in unison as a function of depth are focus of this work.

\subsection{Determination of nuisance covariance from IceCube calibration data\label{sec:NuisCov}}

As is common in experimental particle physics, IceCube uses calibration data to constrain systematic uncertainties that may impact physics analyses.  Given a set of nuisance parameters $\vec{\eta}$, the constraints from calibration data can, under a suitable and testable assumption of Gaussianity, be expressed as a covariance matrix $\Xi$ relative to the best fit point.   The nuisance parameters of interest here the amplitudes and phases of the lowest-frequency modes of $M+$. The covariance matrix in this space encodes the allowed variations in the nuisance parameters, as well as their correlations. 

While ice model calibrations can in principle be performed in many ways, a particularly powerful suite of calibration samples for IceCube are data taken using the LED flashers installed on each Digital Optical Module.  The arrival time distribution and intensity of detected light detected by pairs of flashing and receiving Digital Optical Modules gives a geometry-dependent constraint on absorption and scattering.   These data can be used to derive $\Xi$ for the $M+$ modes of relevance to physics analysis samples.

The covariance matrix construction to describe constraints on the nuisance parameters (for example, properties of the ice) rests on the assumption of a multivariate Gaussian likelihood profile $L$ in the vicinity of the best fit point.  This is to be expected based on the central limit theorem, in most cases.  The covariance matrix then encodes the shape of the likelihood surface around its minimum:
\begin{equation}
L_{\eta_{0},\Xi}(\vec{\eta})=\frac{1}{\sqrt{(2\pi)^{d}det(\Xi)}}exp\left\{ -\frac{1}{2}(\vec{\eta}-\vec{\eta}_{0})^{T}\Xi^{-1}(\vec{\eta}-\vec{\eta}_{0})\right\}.
\end{equation}

We measure $\Xi$ by probing $L$.  The approach we take here is to pick specific directions in which to profile $L$ around the best fit point, in particular in the directions along single nuisance parameter variations, and in fully correlated variations of pairs of nuisance parameters, and find points where the likelihood $L$ has shifted by an amount corresponding to a $1\sigma$ variation in each direction. From these points we extract $\mathcal{M}\equiv \Xi^{-1}$, which can then be inverted to yield the nuisance covariance matrix $\Xi$. The details of the likelihood construction used to compare ice models to calibration data in IceCube is discussed elsewhere~\cite{Refworks:1}.  In the vicinity of the minimum it has been validated that this profile is Gaussian to a very good approximation, in all scanned single-parameter and covariant directions.    
\begin{figure}[t]
\centering\includegraphics[width=\textwidth]{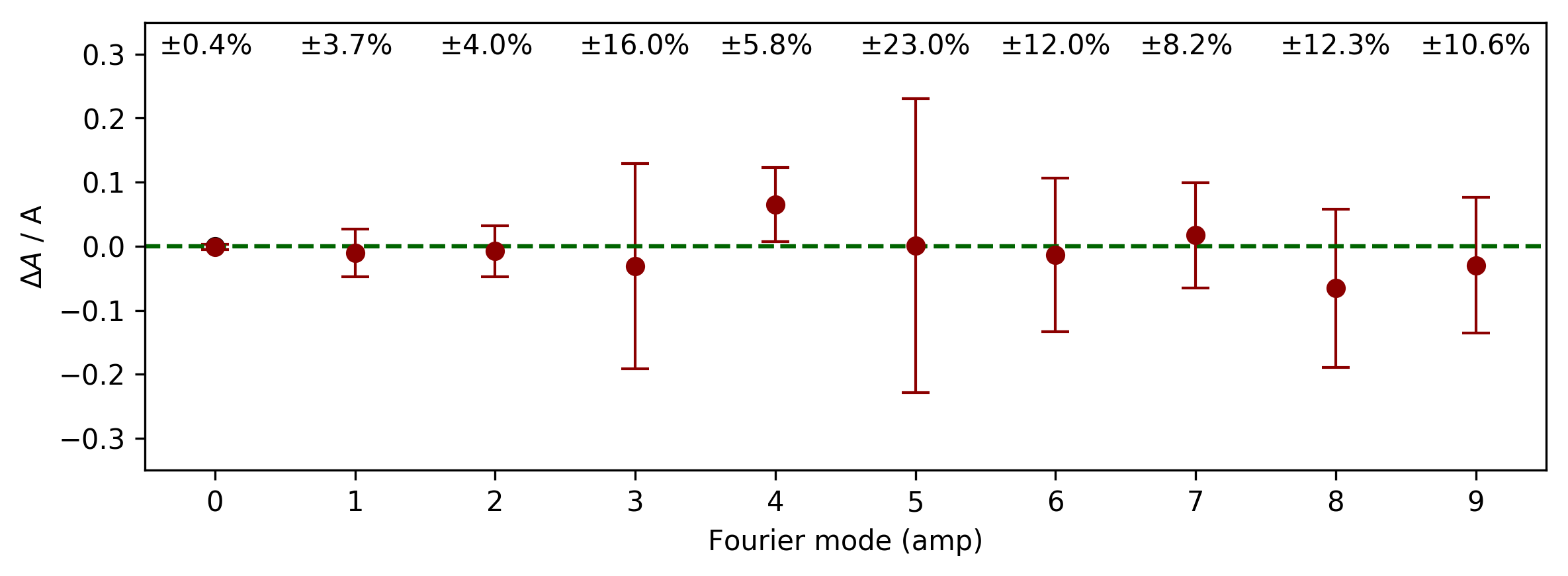}
\centering\includegraphics[width=\textwidth]{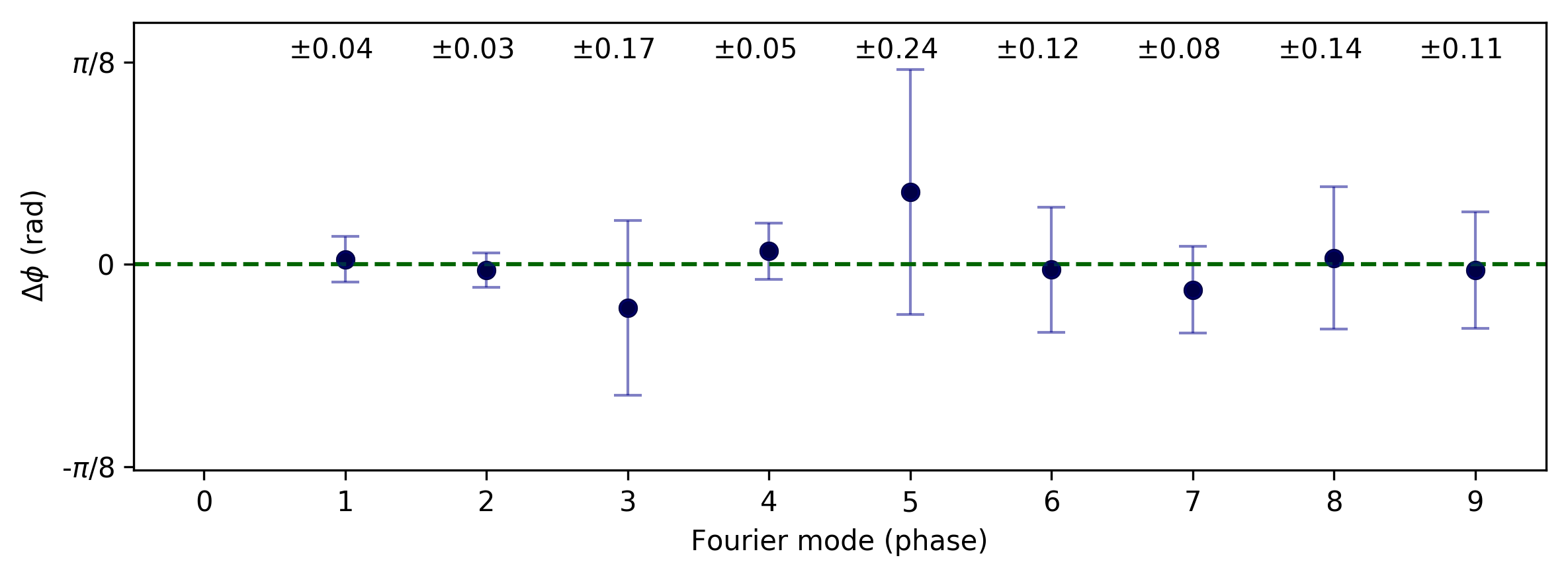}
\includegraphics[width=0.6\textwidth]{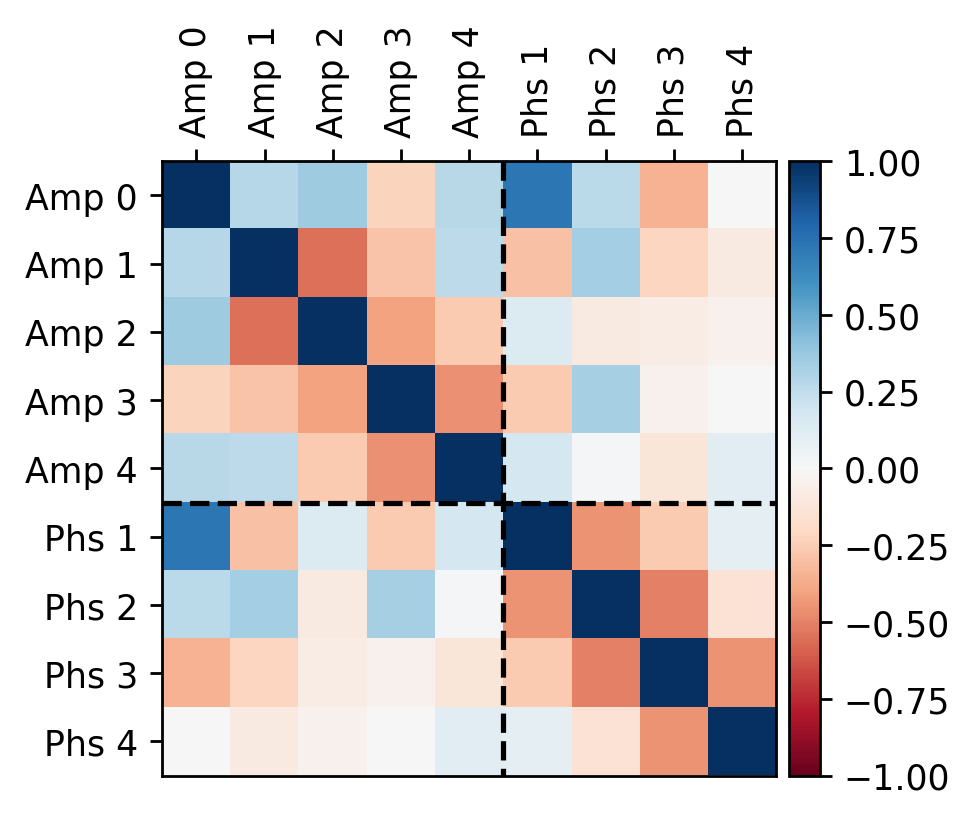}
\caption{Top: Diagonal Fourier parameter constraints derived using flasher calibration data in amplitudes and phases. Bottom: Hessian matrix in nuisance space including correlated parameter widths for a mode 0-4 from flasher calibration data. The dashed lines separate amplitude and phase parameters.}
\label{fig:Constraints}
\end{figure}
The one-sigma points are defined as vectors $\vec{\sigma}$ that may
point in any direction, as:
\begin{equation}
\vec{\sigma}{}^{T}\mathcal{M}\,\vec{\sigma}=\mathcal{M}_{ab}\sigma_{a}\sigma_{b}=1.
\end{equation}
We define two classes of one-sigma vectors $\vec{\sigma}$. First,
for the unit vector directions, $\vec{\sigma}_{unit}^{i}$ for each
$i$. Each vector $\vec{\sigma}^{i}$ with superscript $i$ points in direction
$\hat{i}$, that is:
\begin{equation}
\vec{\sigma}_{unit}^{i}=\sigma_{unit}^{i}\hat{i}\quad\quad[no\,sum\,on\,i],\label{eq:Units}
\end{equation}
or if we include the vector index b explicitly, $
\sigma_{unit,b}^{i}=\sigma_{unit}^{i}\delta_{ib}$ We also define the one-sigma points in the correlated directions. These
are labelled by two indices $\vec{\sigma}_{corr}^{ij}$ , and the vector
$\vec{\sigma}^{ij}$ with subscripts $i$ and $j$ points in a direction
$(\hat{i}+\hat{j})/\sqrt{2}$, that is:
\begin{equation}
\vec{\sigma}_{corr}^{ij}=\sigma_{corr}^{ij}\left(\frac{\hat{i}+\hat{j}}{\sqrt{2}}\right)\quad\quad[no\,sum\,on\,ij]\label{eq:Correlateds}.
\end{equation}
Despite being labelled
by two superscripts, each $\vec{\sigma}_{corr}^{ij}$ is still a
vector with a single vector index, which we can again write as b, as $\sigma_{corr,b}^{ij}=\sigma_{corr}^{ij}\left(\delta_{ib}+\delta_{jb}\right)/\sqrt{2}$.
The unit direction one-sigma
points allow extraction of the diagonal elements of $\mathcal{M}$, via Eq.
\ref{eq:Units}:
\begin{equation}
\mathcal{M}_{ab}\sigma_{unit,a}^{i}\sigma_{unit,b}^{i}=\mathcal{M}_{ab}\delta_{ia}\delta_{ib}\left(\sigma_{unit}^{i}\right)^{2}=1\quad\rightarrow\quad \mathcal{M}_{ii}=\frac{1}{\sigma_{i,unit}^{2}}\quad[no\,sum\,on\,i].
\end{equation}
The correlated directions similarly give the following constraints
on $\mathcal{M}$:
\begin{eqnarray}
1&=&\sum_{ab}\mathcal{M}_{ab}\sigma_{corr,a}^{ij}\sigma_{corr,b}^{ij}\quad\quad[no\,sum\,on\,ij],
\\
&=&\frac{1}{2}\left(\sigma_{corr}^{ij}\right)^{2}\left(\mathcal{M}_{ii}+\mathcal{M}_{jj}+\mathcal{M}_{ij}+\mathcal{M}_{ji}\right).
\end{eqnarray}
Since $\mathcal{M}$ can be defined without loss of generality as a symmetric
matrix, and we have already obtained equations for the diagonal terms, we can invert this expression to find the off-diagonal elements
$\mathcal{M}_{ij}$:
\begin{equation}
\mathcal{M}_{ij}=\frac{1}{\left(\sigma_{corr}^{ij}\right)^{2}}-\frac{1}{2}\left(\frac{1}{\sigma_{i,unit}^{2}}+\frac{1}{\sigma_{j,unit}^{2}}\right).
\end{equation}
Thus with the unit and correlated one-sigma points we can specify the entire
matrix $\mathcal{M}_{ij}$ that defines the multivariate Gaussian, via $\Xi=\mathcal{M}^{-1}$.  The diagonal 1$\sigma$ widths for the leading Fourier amplitudes and phases from IceCube flasher data and the full Hessian matrix $\mathcal{M}$ normalized to these values are shown in  Fig.~\ref{fig:Constraints}. 
\begin{figure}[t]
\centering

\includegraphics[width=0.8\textwidth]{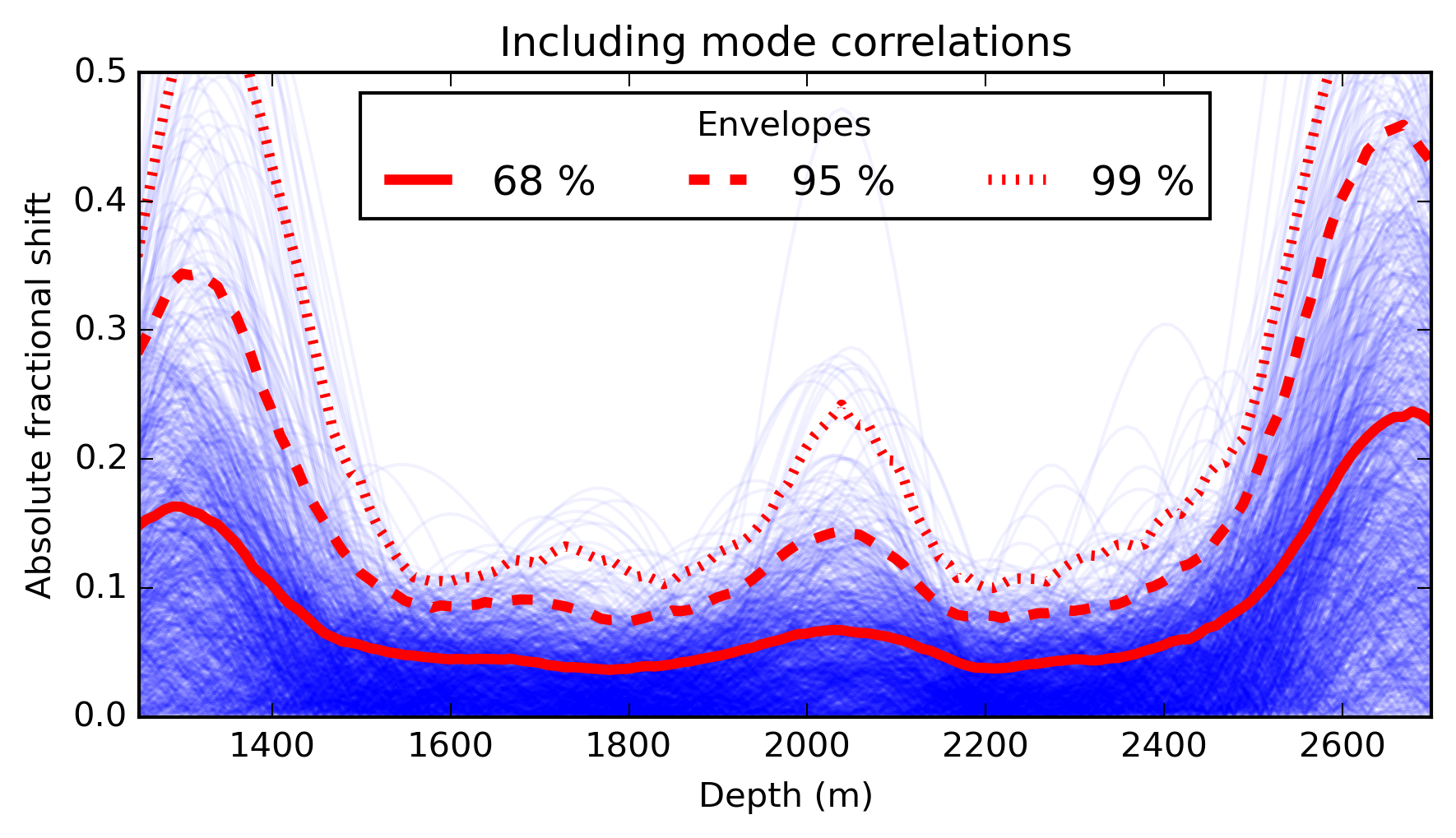}
\includegraphics[width=0.8\textwidth]{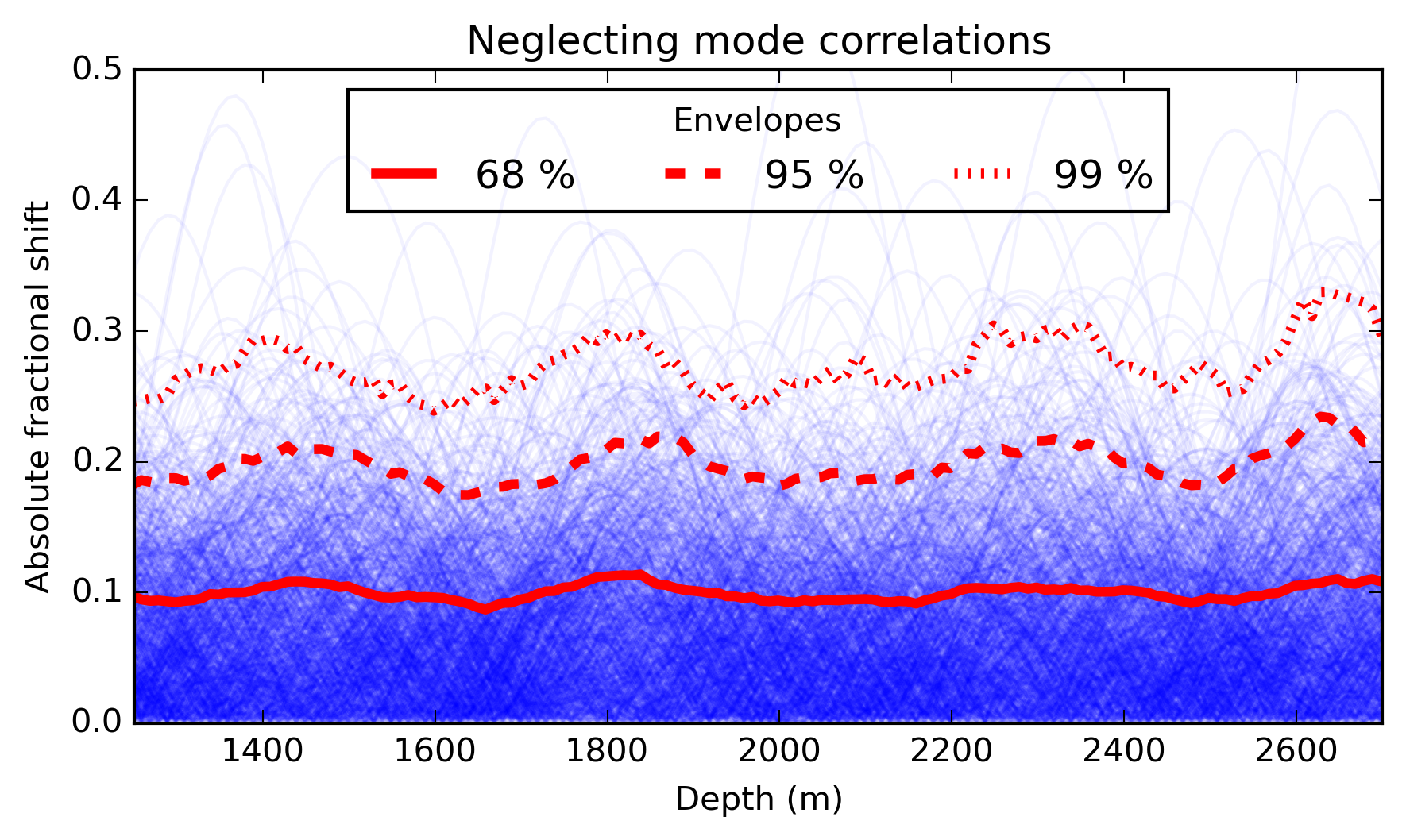}
\caption{Fractional excursions arising from multiple sampling of 5,000 models. The largest excursion are allowed in regions where the properties of the ice are poorly constrained, such as outside the detector and in the dust layer.  Note that this structure is embodied in the correlations between nuisance parameters, rather than their individual spreads - this can be seen by comparing the correlated sampling approach (top) using full covariance matrix with an uncorrelated sampling (bottom).}
\label{fig:env}
\end{figure}

The relevance of the correlation terms is well illustrated by considering the envelope of allowed perturbed models in depth space. Fig.~\ref{fig:env} shows the envelope of allowed correlated variations within the measured covariace matrix, both including off-diagonal correlation terms (top), and treating each mode as independent (bottom).   Red lines mark the edges of frequentist envelopes containing 68\%, 95\% and 99\% of sampled models at each depth.  The emergence of regions of high uncertainty outside of the detector volume, and in the strongly absorbing ``dust layer'' that is known to reside at around -100m in IceCube coordinates, are only manifest when nuisance parameter correlations are properly accounted for. This demonstrates a case where the often overlooked correlations between nuisance parameters constrained by common calibration data are crucial for properly encoding physical features of the underlying detector model.  

The envelope described above was generated with a zero-to-four mode model (9 nuisance  parameters), which limits the sharpness of features in the depth envelope. The choice of a fourth-mode cut-off is motivated in Sec.~\ref{sec:GradientsMEOWs} in the context of the IceCube atmospheric neutrino sample, where modes above the fourth are shown to have a rapidly diminishing effect on the analysis distributions within flasher-constrained uncertainty. Because the size of the envelope in Fig.~\ref{fig:env} depends on the high frequency cutoff, caution should be exercised against over-interpreting as an absolute statement of dust uncertainty vs depth. In principle, very high frequency modes are neither constrained by flasher calibration nor relevant for the atmospheric neutrino sample, and so any envelope so constructed relies on some notion of smoothness or regularization. Our goal here is not to construct a measurement of dust concentration vs. depth, but rather to map the physically important constraints on low frequency modes from calibration data to analysis space, and it is the effect of these modes only that is encoded in Fig.~\ref{fig:env}.

\subsection{Extraction of Fourier nuisance gradients\label{sec:GradientsMEOWs}}


\begin{figure}[t]
\centering
\includegraphics[width=\textwidth]{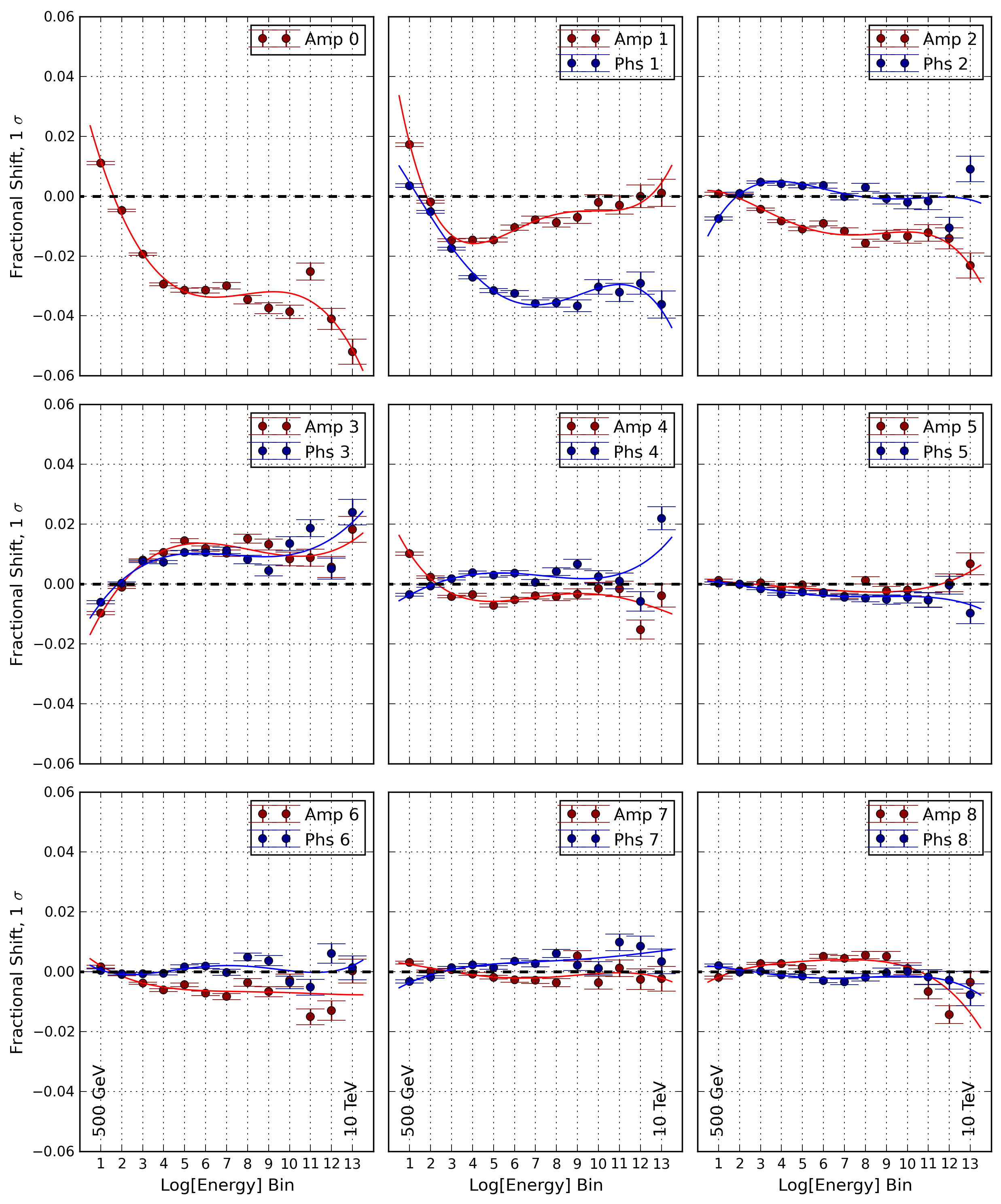}
\caption{Fractional energy gradients split along amplitudes 0 to 8 (red), and phases 1 to 8 (blue). Note the decreasing impact of increasing mode number, verifying the assumption of negligible impact of higher modes.}
\label{fig:E_Grads}
\end{figure}


A SnowStorm ensemble was generated with variations in the first 12 Fourier modes, and processed through photon propagation, detector simulation, Level 1 and Level 2 processing and event selection for atmospheric neutrinos. This processing and selection chain derives from the one described in \cite{Aartsen:2015rwa} with several improvements, and is used for the forthcoming IceCube high energy sterile neutrino analysis. That analysis, which represents the multi-year extension of previous work~\cite{TheIceCube:2016oqi}, selects events between 500~GeV and 10~TeV and compares the energy and zenith distribution of up-going muons against predictions made under different exotic oscillation scenarios.  Hypothetical eV-scale sterile neutrinos, which have been invoked to explain previous short-baseline neutrino oscillation anomalies~\cite{Athanassopoulos:1996jb,Aguilar-Arevalo:2013pmq}, would induce a characteristic disappearance of muon neutrinos due to matter-resonant oscillations. Such a disappearance effect would be observable as a distortion in the reconstructed spectrum, given sufficiently large sterile neutrino mixing.  Accuracy of this analysis relies on a proper description of the effects of systematic uncertainties at the few-percent level over the full spectrum.

Using the cutting based procedure described in Sec.~\ref{sec:SnowStorm}, the vector of gradients $\vec{G}_{\vec{\rho}} \equiv \vec{\nabla}_{\eta}\left[\psi_{\vec{\rho},\vec{\eta}}\right]_{\vec{\eta}=\vec{0}}$ were extracted.  Since existing constraints limit the sterile neutrino parameter space to that where only relatively small perturbations to the atmospheric neutrino spectrum are viable, a single $\vec{G}$ evaluated at $\rho=\vec{0}$ is appropriate.

The first nine gradients for Fourier amplitude and phase parameters as a function of reconstructed energy are shown in Fig.~\ref{fig:E_Grads}.  The scale of effect of perturbing each mode within its $1\sigma$ diagonal uncertainty quickly diminishes with increasing mode number.  Modes beyond approximately mode 4 become irrelevant in terms of their contributions to analysis uncertainty. This motivates restriction of the nuisance covariance and derived analysis covariance to four-mode model. Physically, this represents the principle that the effects of rapid changes in absorption and scattering strength averaging to zero are generally integrated out in the final energy distribution.

The gradients were also examined in zenith space, and found to be negligible there.  The dominant effect of the dust distribution uncertainty is thus demonstrated to be primarily exhibited as distortions to the energy spectrum, and not the zenith distribution of events. The final covariance matrix is thus constructed in energy space only.

Although the trends are clear, the gradients retain some scatter from statistical precision of the SnowStorm ensemble set. In order to produce smooth gradients and avoid artifacts due to Monte Carlo discretization effects, we fit smoothed curves to the extracted fractional gradients.  Empirically, a fourth order polynomial provides an adequate fit in all cases. The best fit curves are shown in Fig.~\ref{fig:E_Grads} as solid lines.  These smoothed gradients are propagated into the analysis covariance matrix.

\subsection{Determination of analysis covariance \label{sec:MEOWSSpace}}

\begin{figure}[t]
\centering
\includegraphics[width=0.7\textwidth]{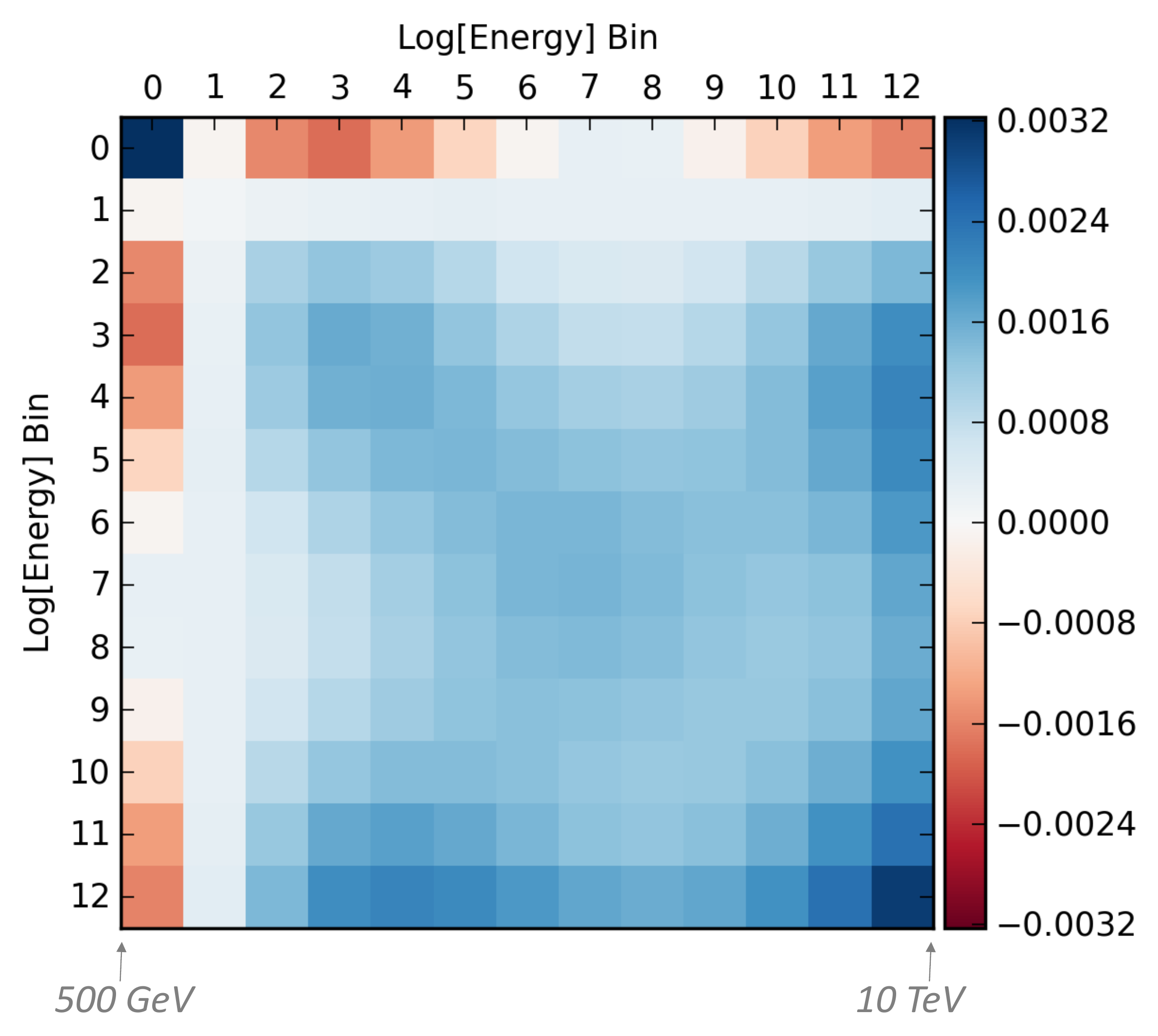}
\caption{ Analysis covariance matrix for application in atmospheric neutrino binned analysis space.}
\label{fig:A_Cov}
\end{figure}

The gradients $\vec{G}$ and the nuisance covariance matrix from calibration data $\Xi$ can finally be combined into an analysis space covariance matrix $\Sigma$, via Eq.~\ref{eq:TheTheorem}.  The smallness of the effect in zenith space allows to concentrate the statistical precision of the sample on achieving the most accurate possible estimate of the energy shape covariance.  The final analysis covariance matrix is shown in Fig.~\ref{fig:A_Cov}.  This matrix represents the depth-dependent dust uncertainty within the atmospheric neutrino sample between 500 GeV and 10 TeV. It incorporates the effects of the physically important modes of a depth-dependent continuous function representing effective dust distribution in the array, properly including effects of correlations between parameters, to yield a fully covariant estimate of uncertainty on the final sample.  Most notably, despite depending on a large number of free parameters, this was achieved using a single Monte Carlo ensemble, via the SnowStorm method.

\subsection{Tests of validity}

Several assumptions are made in the construction of the SnowStorm-derived covariance matrix.  Several of these assumptions can be explicitly tested.  Here we present two examples. 

\begin{figure}[t]
\centering
\includegraphics[width=0.99\textwidth]{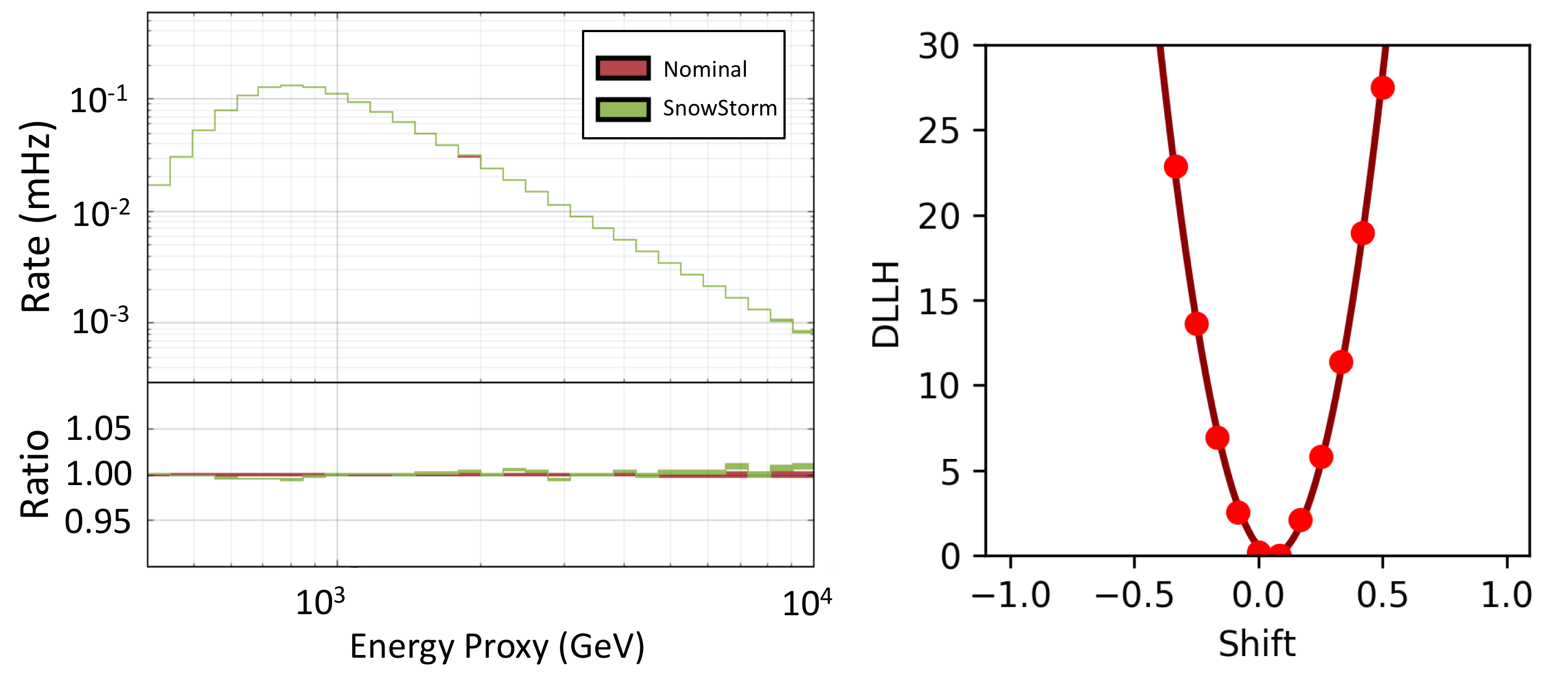}
\caption{Two tests of validity of the SnowStorm method. Left: comparison of integrated SnowStorm ensemble to default Monte Carlo. Lack of substantial difference illustrates that higher order terms can be neglected.  Right: Example delta-log-likelihood (DLLH) profile from calibration data in one of the amplitude directions (mode 4).  All such likelihood profiles are found to be Gaussian within statistical precision.}
\label{fig:CrossChecks}
\end{figure}

First, the principle that the gradients $\vec{G}_{\vec{\rho}} \equiv \vec{\nabla}_{\eta}\left[\psi_{\vec{\rho},\vec{\eta}}\right]_{\vec{\eta}=\vec{0}}$ encapsulate the effects of small variations in the parameters $\eta$ relies on approximate linearity, or equivalently, that the effects of the higher order terms in Eq.~\ref{eq:expansion} are negligible.  Were this not the case, a distribution of the variable of interest made using the full SnowStorm ensemble would not be equivalent to a similar distribution constructed using an unperturbed Monte Carlo set (see Eq.~\ref{eq:TaylorSnow}).  An explicit test of linearity is thus provided by comparing these two distributions, with any non-equivalence indicating the scale of neglected higher-order effects.  A direct comparison of these two distributions for this analysis is shown in Fig.~\ref{fig:CrossChecks}, left, demonstrating agreement at the sub-percent level, at all energies. 

Second, the principle that the constraints on nuisance parameters can be described by a covariance matrix  $\Xi$ requires that the likelihood profile from calibration data is adequately described by a multivariate Gaussian near the best fit.  For this study we found the likelihood from calibration data to be well described by Gaussian profiles in all single- and two-parameter directions, within statistical precision.  Fig.~\ref{fig:CrossChecks}, right shows the delta-log-likelihood distribution between several model perturbations and the best-fit point for variations of the amplitude of mode 4, as an example case.  This is overlaid on a Gaussian profile, demonstrating a high degree of consistency. All other inspected profiles showed a similar level of consistency with the multivariate Gaussian assumption.

\section{Conclusions}
We have presented the SnowStorm method, an approach for efficiently propagating multivariate systematic uncertainties from calibration constraints to analysis space.  Using a perturbative treatment of nuisance parameters, construction of a single Monte Carlo ensemble with randomly distributed nuisance variables allows for construction of predictions given generic nuisance vectors.  The nuisance gradients extracted can then be combined with constraints in the nuisance space derived from calibration data.  The power of this approach lies in the fact that only a single Monte Carlo ensemble is required for treatment of uncertainties describable by an in principle arbitrarily large number of nuisance parameters.

As an example case, an application of the SnowStorm method to IceCube atmospheric neutrino analyses was described. There, a class of bulk ice optical uncertainties related to depth-dependent effective dust concentration within the IceCube array was propagated from LED calibration data to the analysis space of the energy distributions of atmospheric neutrinos.  This application is especially illustrative of the power of this method, since it involves a large number of nuisance parameters that have physically important correlations in their calibration constraints, mapped to analysis covariance via a single Monte Carlo ensemble.  This method is being used by forthcoming high statistics IceCube analyses, and may find wider utility for treating complicated systematic uncertainties, both within IceCube and beyond.

\section*{Acknowledgements}

The IceCube collaboration acknowledges the significant contributions to this manuscript from the University of Texas at Arlington and Massachusetts Institute of Technology groups.  The authors gratefully acknowledge the support from the following agencies and institutions: USA – U.S. National Science Foundation-Office of Polar Programs, U.S. National Science Foundation-Physics Division, Wisconsin Alumni Research Foundation, Center for High Throughput Computing (CHTC) at the University of Wisconsin-Madison, Open Science Grid (OSG), Extreme Science and Engineering Discovery Environment (XSEDE), U.S. Department of Energy-National Energy Research Scientific Computing Center, Particle astrophysics research computing center at the University of Maryland, Institute for Cyber-Enabled Research at Michigan State University, and Astroparticle physics computational facility at Marquette University; Belgium – Funds for Scientific Research (FRS-FNRS and FWO), FWO Odysseus and Big Science programmes, and Belgian Federal Science Policy Office (Belspo); Germany – Bundesministerium für Bildung und Forschung (BMBF), Deutsche Forschungsgemeinschaft (DFG), Helmholtz Alliance for Astroparticle Physics (HAP), Initiative and Networking Fund of the Helmholtz Association, Deutsches Elektronen Synchrotron (DESY), and High Performance Computing cluster of the RWTH Aachen; Sweden – Swedish Research Council, Swedish Polar Research Secretariat, Swedish National Infrastructure for Computing (SNIC), and Knut and Alice Wallenberg Foundation; Australia – Australian Research Council; Canada – Natural Sciences and Engineering Research Council of Canada, Calcul Québec, Compute Ontario, Canada Foundation for Innovation, WestGrid, and Compute Canada; Denmark – Villum Fonden, Danish National Research Foundation (DNRF), Carlsberg Foundation; New Zealand – Marsden Fund; Japan – Japan Society for Promotion of Science (JSPS) and Institute for Global Prominent Research (IGPR) of Chiba University; Korea – National Research Foundation of Korea (NRF); Switzerland – Swiss National Science Foundation (SNSF); United Kingdom – Department of Physics, University of Oxford.


\section*{Appendix: General error propagation and extraction of analysis covariance}

We begin with the most general form of error propagation in a function vector $\vec{f}(\vec{x})$ and proceed via Taylor expansion so that for a small shift in the input parameters $\vec{x}\rightarrow\vec{x}+\delta\vec{x}$ we have:
\begin{equation}
\delta f_{i}=f_{i}(x)+\frac{\partial f_{i}}{\partial x_{j}}\delta x_{j}=J_{ij}\delta x_{j},
\end{equation}
where $J$ is the Jacobian matrix:
\begin{equation}
J_{ij}=\dfrac{df_{i}}{dx_{j}}.
\end{equation}
For the propagation of uncertainties, we seek the covariance $\Sigma_{f}$ of f with respect to its components. We define this as:
\begin{equation}
\Sigma_{f}=\left\langle \left[\vec{f}-E[\vec{f}]\right]\left[\vec{f}-E[\vec{f}]\right]\right\rangle.
\end{equation}
Expanding, we find:
\begin{equation}
\Sigma_{f}=\left\langle f_{i}f_{j}-\bar{f}_{i}f_{j}-f_{i}\bar{f_{j}}+\bar{f_{i}}\bar{f}_{j}\right\rangle =\left\langle f_{i}f_{j}\right\rangle -\left\langle \bar{f}_{i}f_{j}\right\rangle -\left\langle f_{i}\bar{f_{j}}\right\rangle +\left\langle \bar{f_{i}}\bar{f}_{j}\right\rangle,
\end{equation}
with all three of the last terms reducing to the same expression:
\begin{equation}
\left\langle \bar{f}_{i}f_{j}\right\rangle =\left\langle f_{i}\bar{f_{j}}\right\rangle =\left\langle \bar{f_{i}}\bar{f}_{j}\right\rangle =\bar{f_{i}}\bar{f}_{j},
\end{equation}
and so:
\begin{equation}
\Sigma_{f}=\left\langle f_{i}f_{j}\right\rangle -\bar{f_{i}}\bar{f}_{j}.
\end{equation}
If the variability in $f$ around its mean is due to small perturbations
in x we can write $f=\bar{f}+J.\delta x$ for randomly fluctuating
$\delta x$ and so:
\begin{equation}
\Sigma_{f}=\left\langle \left(\bar{f}_{i}+J_{ik}\delta x_{k}\right)\left(\bar{f}_{j}+J_{jl}\delta x_{l}\right)\right\rangle -\bar{f_{i}}\bar{f}_{j},
\end{equation}
\begin{equation}
\Sigma_{f}=\bar{f_{i}}\bar{f}_{j}+\bar{f}_{i}\mathcal{J}_{ik}\left\langle \delta x_{k}\right\rangle +\bar{f}_{j}J_{jl}\left\langle \delta x_{l}\right\rangle -\bar{f_{i}}\bar{f}_{j}+J_{ik}J_{jl}\left\langle \delta x_{k}\delta x_{l}\right\rangle.
\end{equation}
Since the $\delta x_{i}$ all fluctuate around 0, we know:
\begin{equation}
\left\langle \delta x_{i}\right\rangle =0\quad\forall\quad i.
\end{equation}
We can also identify the last ensemble average as the covariance in
x:
\begin{equation}
\left\langle \delta x_{k}\delta x_{l}\right\rangle \equiv\left[\Sigma_{x}\right]_{kl},
\end{equation}
leading us to the central identity, as required:
\begin{equation}
\Sigma_{f}=J\Sigma_{x}J^{T}.
\end{equation}
In our case of interest we associate $f\rightarrow\psi_{\vec{\rho},\vec{\eta}}$ and $x\rightarrow\vec{\eta}$. Then $\Sigma_f$ is the covariance matrix in analysis space, $\Sigma_f\rightarrow\Sigma$, and $\Sigma_x$ is the covariance matrix in nuisance space, $\Sigma_x\rightarrow\Xi$. Then $J_{\alpha,i}=\partial  \psi_{\vec{\rho},\vec{\eta}} (E_\alpha) / \partial \eta_i = G_{\vec{\rho},i;\alpha}$, proving the theorem
\begin{equation}
\Sigma_{\alpha,\beta}=G_{i;\alpha} \Xi_{ij} G_{j;\beta}.
\end{equation}
\bibliographystyle{unsrt}
\bibliography{references}

\end{document}